\documentclass[pra,nofootinbib]{revtex4}

\usepackage{graphicx}
\usepackage{amssymb}
\usepackage{amsmath}
\usepackage{amsfonts}
\usepackage{epstopdf}
\usepackage{epsfig}
\usepackage{wrapfig}

\begin{document}

\title{A quantum open system model of molecular battery charged by excitons}
\author{Robert Alicki}
\affiliation{International Centre for Theory
of Quantum Technologies (ICTQT), University of Gda\'nsk, 80-308, Gda\'nsk, Poland }

\begin{abstract}
The analytically tractable model employing  Quantum Markovian Master Equations, derived by weak coupling procedure and satisfying complete positivity, is proposed to describe a model of molecular battery charged by a non-equilibrium excitonic reservoir. The excitons are produced by non-equilibrium processes involving, e.g. light absorption and chemical reactions. Various  relations concerning the efficiency of the involved processes of energy transfer and the stability of battery are discussed. The model can be treated as an initial step in applications of mathematically sound version of Quantum Theory of Open Systems to complex processes of energy transfer in biological system and man-made devices based on organic materials.

\end{abstract}

\maketitle
\section{Introduction}
Energy transduction processes on the molecular level are responsible for functioning of living organisms and operation of various man-made devices like photovoltaic, thermoelectric or chemical cells.  The ultimate microscopic theory of these phenomena should be quantum and derivable from the first principles, i.e. from the fully quantum Hamiltonian models. The natural mathematical formalism to achieve this task is provided by the Quantum Theory of Open Systems \cite{Alicki:07},\cite{Breuer},\cite{Huelga}, in particular by the theory of Quantum Markovian Master Equations (QMME). 
Those  equations of motion for the reduced density matrices of the open systems are obtained by elimination of the environmental degrees of freedom. From the plethora of various approaches and mathematical tools we select the method introduced by Davies \cite{Davies} combining the weak coupling (Born), Markovian and secular approximations. The obtained QMMEs satisfy the requirement of complete positivity,\cite{GKS},\cite{Lindblad} possess a clear mathematical structure and are consistent with thermodynamics \cite{SpohnLeb},\cite{Alicki:79},\cite{QT-review}. In particular the last property is important because the universal laws of thermodynamics provide the independent validity control of the often involved approximation schemes.
\par
The aim of this paper is to present an analytically tractable model based on QMMEs which are derivable from the full Hamiltonian dynamics and describe a certain sequence of energy transduction processes. Namely, in the first step the electronic system (solid state or organic material) is weakly coupled to a stationary but non-equilibrium environment. The thermal and/or chemical energy supplied by the reservoir is used to produce a gas of excitons which can be treated as a ``chemical fuel" used, for example, to power various types of engines (e.g. photovoltaic cells, photosynthetic reaction centers). The models of such engines based on QMMEs and employing self-oscillation mechanisms were studied for example in \cite{Solar}, \cite{Universal}. Here, the different process of energy transduction is discussed --  charging of ``molecular battery" by non-equilibrium exciton gas. The presented model of molecular battery, involving a single ``reaction coordinate" and two electronic states, can be seen, for example,  as a toy model of ATP/ADP complex while the whole discussed sequence of  energy transfers can mimic the complicated  processes in photosynthetic light harvesting systems. Finally, the process of spontaneous discharging of molecular battery at ambient temperature is studied in order to estimate stability of battery.
\par
The main advantage of the applied version of QMMEs is their particular mathematical structure with clear thermodynamical interpretation. This allows to derive various relations characterizing the thermodynamical efficiency of energy transduction processes and the corresponding trade-offs between physical parameters of the model.
\par
For the reader's convenience the Appendix with a brief derivation of QMME is added. An intermediate step, called recently \emph{refined weak coupling} \cite{Rivas} and derived much earlier in \cite{Alicki:89} as  completely positive non-Markovian Born approximation, is used. This formalism provides a mathematically consistent and apparently numerically accurate interpolation between a slow highly non-Markovian short-time dynamics and long-time Markovian one governed by QMME. 

\section{Exciton factory}

Exciton is a neutral elementary excitation of an electronic system  composed of electron and hole forming a bound state. In semiconductors the binding energy is usually small with respect to ambient thermal energy $k_BT$ and hence one can treat exciton as an essentially free electron excited from the valence to conduction band leaving behind a hole in the valence band. The bands are assumed to be separated by the gap $E_g >> k_BT$ and the valence band is completely filled at zero temperature. Therefore each excited electron can be identified with the created exciton. For organic materials the HOMO (highest occupied molecular orbital) level replaces the valence band maximum  and the LUMO (lowest unoccupied molecular orbital) level  the conduction band minimum. 
Here, binding energy of the exciton is higher but again the picture of independent electrons with a certain self-consistent free Hamiltonian provides a reasonable approximation.
\par
The exciton production is the first step of the discussed energy transfer process involving also charging of molecular battery and its controlled or spontaneous discharging, as presented schematically on Fig.1.

\begin{figure} [t]
	\begin{center}
		\includegraphics[width=1.0 \textwidth]{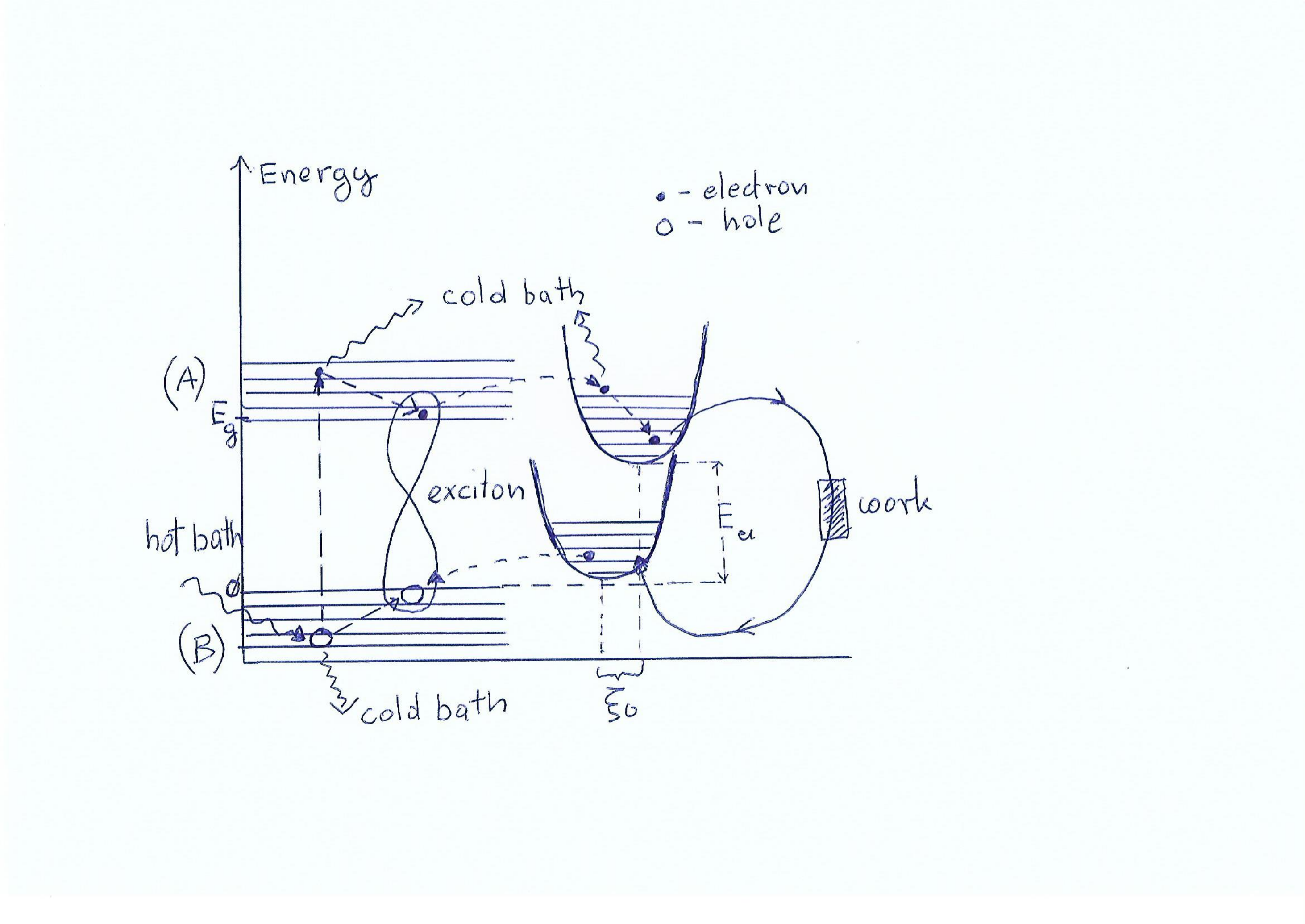}
	\end{center}
\caption{\small Schematic representation of the operation of  molecular battery: I) creation of exciton at the expense of heat or chemical energy of the hot bath, II) electron-hole thermalization to the cold bath temperature, III) exciton recombination charging the battery and  accompanied by thermalization of reaction coordinate, IV) work extraction or spontaneous discharging.}
\end{figure}

\par
To describe creation of excitons  we use a fully quantum mechanical model of a system with relevant degrees of freedom corresponding to electrons distributed in two bands $A$ (higher) and $B$ (lower), separated by the energy gap $E_g$, and interacting with a non-equilibrium environment. The environment consists of two components: the cold, typically phonon bath at the ambient temperature $T$  and the driving hot bath which can be in a non-equilibrium but stationary state characterized by the local temperature $T[\epsilon]$ depending on the energy scale $\epsilon$ \cite{localT} and  fixed chemical potentials $\mu_j$ corresponding to different chemical species  being subjected to a chemical reaction. The example of such a bath is a stationary but generally non-thermal light source parametrized by photon populations $n(\omega) = [e^{\hbar\omega/k_BT[\hbar\omega]}-1]^{-1}$. A chemical bath can  produce a single  exciton at the expense  of  the free energy $\Delta g $  per single reaction step given by 
\begin{equation}
\Delta g = -\sum_{j=1}^K \nu_j\mu_j ,
\label{free_gibbs}
\end{equation}
where $\nu_j$ are \emph{stoichiometric coefficients}, negative for reactants and positive for reaction products (see Appendix). 

\subsection{Model Hamiltonians}
 
The electrons distributed in two bands $A$ and $B$ are described by two sets of  fermionic annihilation and  creation operators $a_{k}$, $a^{\dagger}_{k}$ and $b_{l}$, $b^{\dagger}_{l}$, respectively.
The electrons are treated as  non-interacting fermions moving in a self-consistent potential with the Hamiltonian  
\begin{equation}
H_{0} =  \sum_{k} E_{a}(k) a^{\dagger}_{k} a_{k} +  \sum_{\ell} E_{b}(\ell) b^{\dagger}_{\ell} b_{\ell}. 
\label{ham_electrons}
\end{equation}
Here, $k$ and $\ell$ is a short-hand notation for a set of quantum numbers characterizing a single-electron energy eigenstate. Notice, that the spatial structure of the sample is hidden in the corresponding wave functions $\phi_k (\mathbf{r})$, $\phi_{\ell} (\mathbf{r})$ written in the position representation. The often used semi-classical picture of ``band bending" at the interfaces between different materials is not consistent with the fully quantum description.

We can assume that the energy gap $E_g$ is higher than the typical Debye energy of phonons. Then, the  cold phonon bath can cause only \emph{intraband transitions} and the corresponding interaction Hamiltonian can be written as 
\begin{equation}
H_{intra} =  \sum_{kk'}  a^{\dagger}_{k} a_{k'}\otimes R^{(a)} _{kk'}+  \sum_{\ell\ell'}  b^{\dagger}_{\ell} b_{\ell'}\otimes R^{(b)} _{\ell\ell'} 
\label{ham_intra}
\end{equation}
Here, $R^{(a)} _{kk'}, R^{(b)} _{\ell\ell'}$ are hermitian matrices consisting of operators acting on the Fock space of phonons. The detailed form of those operators is not relevant for the further derivations.

Because the intraband transitions are usually dominated by the interaction with cold phonon bath we can assume that  the driving hot bath contributes to \emph{interband transitions} only. In this case the interaction Hamiltonian reads
\begin{equation}
H_{inter} =  \sum_{k\ell} \bigl( a^{\dagger}_{k} b_{\ell} +  b^{\dagger}_{\ell} a_{k}\bigr)\otimes R^{(ab)} _{k\ell}, 
\label{ham_inter}
\end{equation}
where, again, the exact form of bath operators is not relevant.

We are interested in the situation when at zero temperature the lower band $B$ is completely filled, hence the operator $a^{\dagger}_{k} b_{\ell}$  ($b^{\dagger}_{\ell} a_{k}$) creates (annihilates) an electron-hole pair (exciton) from the ground state of electronic systems.

\subsection{Quantum Markovian Master Equation}

The derivations of QMME for the similar models have been presented in previous papers (e.g. \cite{Solar}, \cite{Universal}), therefore we reproduce here only the final results and discuss their physical interpretation. However, for reader's convenience the basis steps of the derivation for a general QMME are presented in the Appendix together with selected properties of them.
\par
The  QMME for the reduced density matrix $\rho(t)$ acting on the fermionic Fock space of electrons reads
\begin{equation} 
\frac{d}{dt} \rho = -\frac{i}{\hbar}[H_{0} , \rho] + \mathcal{L}  \rho ,
\label{MME}
\end{equation}
where $H_{0}$ is the Hamiltonian given by \eqref{ham_electrons} with renormalized single-electron energies including lowest order corrections
due to the interaction with the baths in the self-consistent approximation. The irreversible processes are governed by the generator $\mathcal{L} $ with the following structure 
\begin{equation} 
\mathcal{L}  = \mathcal{L}^{intra} + \mathcal{L}^{inter} ,
\label{L_gen}
\end{equation}
\begin{equation} 
\mathcal{L}^{intra} = \sum_{\{kk'\}}\mathcal{L}^{(a)}_{kk'}  + \sum_{\{\ell\ell'\}}\mathcal{L}^{(b)}_{\ell\ell'} ,
\label{L_intra} 
\end{equation}
\begin{equation} 
\mathcal{L}^{inter}  = \sum_{\{k\ell\}}\mathcal{L}^{(ab)}_{k\ell}  ,
\label{L_inter} 
\end{equation}
\begin{align}
		&\mathcal{L}^{(a)}_{kk'}\rho = \frac{1}{2}\Gamma^{(a)}_{kk'}\Bigl([ a_{k} a_{k'}^{\dagger} \, \rho ,\,a_{k'} a_{k}^{\dagger} ]+ [ a_{k} a_{k'}^{\dagger} , \, \rho \,a_{k'} a_{k}^{\dagger} ]\nonumber \\
		&+ e^{-(E_a(k) -E_a(k')/k_BT} \bigl([ a_{k}^{\dagger} a_{k'} \, \rho , \,  a_{k'}^{\dagger} a_{k} ] + [ a_{k}^{\dagger} a_{k'}, \, \rho  \,  a_{k'}^{\dagger} a_{k} ] \bigr)\Bigr), 
		\label{ME_intraA}
	\end{align}
\begin{align}
		&\mathcal{L}^{(b)}_{\ell\ell'}\rho = \frac{1}{2}\Gamma^{(b)}_{\ell\ell'}\Bigl( [b_{\ell} b_{\ell'}^{\dagger} \, \rho , \,b_{\ell'} b_{\ell}^{\dagger}] + [b_{\ell} b_{\ell'}^{\dagger}, \, \rho  \, b_{\ell'} b_{\ell}^{\dagger}]\nonumber \\
		&+ e^{-(E_b(\ell) -E_b(\ell')/k_BT }\bigl( [b_{\ell}^{\dagger} b_{\ell'} \, \rho ,\,  b_{\ell'}^{\dagger} b_{\ell}] + [b_{\ell}^{\dagger} b_{\ell'}, \, \rho \,  b_{\ell'}^{\dagger} b_{\ell}]\bigr)\Bigr), 
		\label{ME_intraB}
	\end{align}
\begin{align}
		&\mathcal{L}^{(ab)}_{k\ell}\rho = \frac{1}{2}\gamma_{k\ell}\Bigl( [a_{k} b_{\ell}^{\dagger} \, \rho ,\,b_{\ell} a_{k}^{\dagger}] + [a_{k} b_{\ell}^{\dagger}, \, \rho \,b_{\ell} a_{k}^{\dagger}]\nonumber \\
		&+ e^{-(\epsilon_{kl} - \Delta g)/k_BT[\epsilon_{kl}]} \bigl( [a_{k}^{\dagger} b_{\ell} \, \rho ,\,  b_{\ell}^{\dagger} a_{k}] + [a_{k}^{\dagger} b_{\ell} , \, \rho \,  b_{\ell}^{\dagger} a_{k}] \bigr)\Bigr). 
		\label{ME_inter}
	\end{align}
The summations in the formulas of above are performed over the sets of indices corresponding to positive energy differences between pairs of states: $ \{kk'\} = \{kk';E_a(k) -E_a(k')\geq 0\}, \{\ell\ell'\}= \{\ell\ell';E_b(\ell) -E_b(\ell')\geq 0\}, \{k\ell\}=  \{k\ell;E_{a}(k) - E_{b}(\ell)> \geq 0\}$, $\epsilon_{kl} =( E_{a}(k) - E_{b}(\ell))$.
\par
Excitons are created by the operators $a_{k}^{\dagger} b_{\ell}$ and annihilated by $b_{\ell}^{\dagger} a_{k}$.
As explained in the Appendix the Boltzmann factor in \eqref{ME_inter} contains the contribution $\Delta g$ to energy difference which accounts for chemical energy provided by the hot bath.
\par
The term \eqref{ME_intraA} (\eqref{ME_intraB}) describes the following  intraband processes:\\
a) electron relaxation from the state $k$  ($l$) to the state $k'$  ($l'$ ) accompanied by a  positive energy release to the heat bath at the ambient temperature , \\
b) the inverse process of electron transfer from the lower energy state to the higher energy one  with the probability suppressed by the Boltzmann factor.\\
\par
The  term \eqref{ME_inter}   describes the following interband processes:\\
c) electron transfer from the higher energy state $k$ in the  band $A$ to the lower energy state $l$ in the band $B$,\\
d) the inverse process of electron transfer  with the probability controlled by  suitable Boltzmann factors including energy dependent effective temperature $T[\cdot]$ and the  chemical energy $\Delta g$.
\par
In principle all relaxation parameters can be derived from the underlying Hamiltonian models with properly chosen stationary states of the baths. In practice phenomenological values obtained e.g. from the measurements of light absorption and electron thermalisation rates can be used.

\subsection{Stationary state of electronic system}

The  analysis of the stationary state for the QMME \eqref{MME} - \eqref{ME_inter} is based on the following facts:\\
i) the intraband thermalization processes governed by \eqref{ME_intraA}, \eqref{ME_intraB} preserve the number of electrons in each band and for a fixed initial numbers
of electrons drive the system into a product of two canonical Gibbs states  at the temperature $T$,\\
ii) the interband transitions described by \eqref{ME_inter} preserve the total number of electrons  but modify the relative occupation of both band,\\
iii) because the number of electrons  in a macroscopic system is very large we can use the equivalence
of canonical and grand canonical ensembles to derive the convenient approximate form of the stationary state for  $\mathcal{L} $.\\

\par
Taking i)-iv) into account one concludes that the approximate stationary state of the generator  $\mathcal{L} $ has a form of the product of  grand canonical ensembles at the ambient temperature $T$, but with two different electrochemical potentials  $\mu_a , \mu_b$
\begin{equation}
\bar{\rho}[\mu_a , \mu_b] = {Z}^{-1} \exp\Bigl\{ -\frac{1}{k_B T} \Bigl[\sum_{k}  \bigl(E_a(k) -\mu_a\bigr) a^{\dagger}_{k}a_{k} + 
\sum_{l} \bigl(E_b(l)-\mu_b\bigr) b^{\dagger}_{l}b_{l}\Bigr]\Bigr\} . 
\label{grand}
\end{equation}
The states $\bar{\rho}[\mu_a , \mu_b] \equiv \bar{\rho} $ are stationary for the generator $\mathcal{L}^{intra}$ for any choice of chemical potentials. Even their restrictions to subspaces of the Fock space with fixed electron numbers are stationary. To find the best approximation to the stationary state with respect to the full dynamics given by QMME \eqref{MME} we have to minimize the trace norm of $\mathcal{L}^{inter} \bar{\rho}$. This allows to obtain the optimal parameter 
\begin{equation}
\Delta\mu = \mu_a - \mu_b, 
\label{Dmu}
\end{equation}
while the chemical potential $\mu_b$ is fixed by the averaged electron number in the system.
A series of simple computations yields  the following expression for $\mathcal{L}^{inter} \bar{\rho}$
\begin{equation}
		\mathcal{L}^{inter} \bar{\rho} = 
\sum_{k,\ell}\gamma_{k\ell}\Bigl\{\bigl[e^{-X_{k\ell}} - e^{-Y_{k\ell}} \bigr]\bar{\rho}\,a_{k}a_{k}^{\dagger} b_{\ell}^{\dagger}b_{\ell}
+ \bigl[e^{X_{k\ell}} e^{-Y_{k\ell}}-1 \bigr] \bar{\rho}\,
	 a_{k}^{\dagger}a_{k} b_{\ell}b_{\ell}^{\dagger}\Bigr\} , 
\label{Lstat}
\end{equation}
where
\begin{equation}
X_{k\ell} = \beta(\epsilon_{k\ell} - \Delta\mu), \quad Y_{k\ell} = \beta[\epsilon_{k\ell}](\epsilon_{k\ell} -\Delta g).
\label{XY}
\end{equation}
Introducing the Fermi-Dirac distributions
\begin{equation}
f_a(k) = \frac{1}{e^{\beta(E_a(k) - \mu_a)}+1} , \quad f_b(\ell) = \frac{1}{e^{\beta(E_b(\ell) - \mu_b)}+1} 
\label{FD}
\end{equation}
one obtains the following estimation of the trace norm 
\begin{equation}
\|\mathcal{L}^{inter} \bar{\rho}\|_1 \leq \sum_{k,\ell}\gamma_{k\ell} |1- e^{X_{k\ell}-Y_{k\ell}}|
\Bigl\{ e^{-X_{k\ell}} [1- f_a(k)]f_b(\ell) + f_a(k)[1- f_b(\ell)]\Bigr\}
\label{Lstat_norm}
\end{equation}
The trace norm is small if for those energy differences $\epsilon_{k\ell} $ which enter the RHS of \eqref{Lstat_norm} with  the substantial weights, the approximate equality holds  
\begin{equation}
X_{k\ell} \simeq Y_{k\ell}
\end{equation}
what implies
\begin{equation}
\Delta\mu \simeq \Bigl( 1 - \frac{T}{T[\epsilon_{k\ell}]}\Bigr)\epsilon_{k\ell} +  \frac{T}{T[\epsilon_{k\ell}]}\Delta g.
\label{Delta}
\end{equation}
Remembering that $\epsilon_{k\ell} \geq E_g >> k_B T$ and  that the intraband thermalization is much faster than the interband one we expect that only the energy differences close to the gap, i.e. $\epsilon_{k\ell} \simeq E_g +\mathcal{O}( k_B T)$ matter. Hence, in the formula \eqref{Delta} $\epsilon_{k\ell}$ can be replaced by the effective, weakly dependent on the temperature, energy gap $\bar{E}_g \simeq E_g +\mathcal{O}( k_B T)$. As a consequence the final formula for the chemical potential  of the  non-equilibrium excitonic bath fed  by the external temperature gradient and chemical reactions reads
\begin{equation}
\Delta\mu =\Bigl( 1 - \frac{T}{T[\bar{E}_g ]}\Bigr)\bar{E}_g  +  \frac{T}{T[\bar{E}_g ]}\Delta g.
\label{exciton}
\end{equation}
This relation appears in various contexts, for example describes the open circuit voltage ($eV_{oc} \equiv\mu_a - \mu_b$)  in  photovoltaic, thermoelectric and chemical cells \cite{hot},\cite{Universal}. Its thermodynamical meaning is quite clear. The chemical energy of an exciton $\Delta\mu$ can be in principle completely converted into work by exciton recombination in a certain external circuit attached to the excitonic reservoir. The formula shows that if the exciton is created by thermal energy the heat - chemical potential conversion efficiency is limited by the Carnot bound while in the case of isothermal conditions ($T = T[\bar{E}_g ]$) the whole energy released in the chemical reaction can be, in principle, transferred into work ($\Delta\mu = \Delta g$).
\section{Molecular battery}
We consider a model of a molecule with two relevant electronic states $|0\rangle , |1\rangle$ , a ground state and the first excited state, respectively, separated by the energy gap $E_{el}$. The other relevant degree of freedom can be called \emph{reaction coordinate} and is modeled by the harmonic oscillator with a single, independent of the electronic state, frequency $\omega_0$ and the annihilation and creation operators $A , A^{\dagger}$ ( a similar model of a ``quantum switch" has been studied in \cite{Alicki:2013}). It is not difficult to generalize this model to many electronic states with different frequencies for the reaction coordinate at the price of much more complicated formulas.
\par
The  Hamiltonian of the model which is illustrated on  Fig.1 reads
\begin{equation}
H_m = \hbar\omega_0 (A^{\dagger} - \xi_0|1\rangle\langle 1|)(A - \xi_0|1\rangle\langle 1|) + E_{el}|1\rangle\langle 1|,
\label{ham_TLSQB}
\end{equation}
where $\xi_0$ describes the dimensionless displacement of the minimum of the potential curve for excited electronic state with respect to the ground state and is equal to the square root of the Huang-Rhys factor. 

It is convenient to introduce the \emph{polaron transformation} in the form of the following unitary operator
\begin{equation}
U = e^{\xi_0( A^{\dagger} - A)|1\rangle\langle 1|},
\label{dress1}
\end{equation}
which yields the new set of operators for the harmonic oscillator part and a new form of the Hamiltonian
\begin{equation}
A \mapsto B = U A U^{\dagger} = A - \xi_0 |1\rangle\langle 1|, \quad H_{m} = \hbar\omega_0 B^{\dagger} B + E_{el} |1\rangle\langle 1|.
\label{dress2}
\end{equation}
For electronic states
\begin{equation}
|0\rangle\langle 0| \mapsto  |0\rangle\langle 0|, \quad |1\rangle\langle 1| \mapsto  |0\rangle\langle 0|, \quad
|1\rangle\langle 0| \mapsto E_{10} = e^{\xi_0( B^{\dagger}- B)}|1\rangle\langle 0| 
\label{dress3}
\end{equation}
We treat the molecule as an open system subject to the weak interaction with the stationary environment including also the other molecular degrees of freedom. The lowest order interaction Hamiltonians can be chosen as  
\begin{eqnarray}
\label{Ham_int_1}
{H}^{(1)}_{int} &=& (A + A^{\dagger})\otimes {F_1} = (B + B^{\dagger} - 2\xi_0|1\rangle\langle 1| ) \otimes{F}_1,\\
\label{Ham_int_2}
{H}^{(2)}_{int} &=&  |1\rangle\langle 1|\otimes{F}_2  ,
 \\
{H}^{(3)}_{int} &=& (|1\rangle\langle 0|+ |1\rangle\langle 0|) \otimes{F}_3 = \bigl[e^{-\xi_0( B- B^{\dagger})} E_{10}
 + h.c.\bigr]\otimes{F}_3
\label{Ham_int_3}
\end{eqnarray}
where $F_{j}, j = 1, 2, 2$ are statistically independent environment observables. 

The interaction Hamiltonian ${H}^{(1)}_{int}$ is used to account for the thermalization process of the reaction coordinate to the ambient temperature. The second Hamiltonian ${H}^{(2)}_{int}$ commutes with the molecular Hamiltonian, hence describes various pure decoherence processes of electronic states. The last Hamiltonian ${H}^{(3)}_{int}$ governs transitions between electronic states which in our model are dominated by the interaction with the non-equilibrium excitonic reservoir discussed in the previous Section.
\par
In order to construct the proper QMME using again the weak coupling approach we have to calculate the Heisenberg dynamics of relevant molecular operators entering the interaction Hamiltonians \eqref{Ham_int_1} -- \eqref{Ham_int_3}
\begin{eqnarray}
\label{Fourier_1}
e^{\frac{i}{\hbar}{H}_m t}(A + A^{\dagger}) e^{-\frac{i}{\hbar}{H}_m t}  &=& e^{-i\omega_0 t} B + e^{i\omega_0 t}B^{\dagger} 
- 2\xi_0 |1\rangle\langle1|,
\\
\label{Fourier_2}
e^{\frac{i}{\hbar}{H}_m t} |1\rangle\langle 0| e^{-\frac{i}{\hbar}{H}_m t}  &=&  W(\xi_0) W^{\dagger}( e^{i\omega_0 t}\xi_0) 
|1\rangle\langle 0| ,
\end{eqnarray}
where $W(\alpha)$ denotes the Weyl operator
\begin{equation}
W(\alpha) = e^{(\alpha A^{\dagger} - \bar{\alpha} A)} = e^{-|\alpha|^2/2}e^{\alpha A^{\dagger}} e^{ - \bar{\alpha} A} .
\label{weyl}
\end{equation}

The transition operators entering the QMME for the reduced density matrix of the molecule are obtained from the decomposition of the RHS of eqs.\eqref{Fourier_1}, \eqref{Fourier_2} into terms oscillating with relevant Bohr frequencies. 

\subsection{Thermalization of reaction coordinate}

To simplify the analysis of the molecular dynamics we consider first only the thermalization to ambient temperature cause by the coupling of the type \eqref{Ham_int_1}, \eqref{Ham_int_2}. The QMME is constructed using the decomposition into oscillating terms 
\eqref{Fourier_1} and coupling spectra for the thermal bath at the temperature $T$
\begin{equation}
G_j(\omega)= \int_{-\infty}^{+\infty} e^{-i\omega t}\langle \hat{F}_j(t)\hat{F}_j\rangle_{T}\, dt \geq 0 ,\quad j=1,2
\label{spectral_bath_T}
\end{equation}
where $\langle \cdot \rangle_{T}$ denotes the average with respect to Gibbs state of the bath. The following Kubo-Martin-Schwinger (KMS) relation holds
\begin{equation}
G_j(-\omega)= e^{-\hbar\omega/k_BT} G_j(\omega).
\label{KMS}
\end{equation}
Standard derivation procedure leads to the following QMME equation for the reduced density matrix of the molecule
\begin{eqnarray}
\frac{d}{dt}\rho = -\frac{i}{\hbar}[{H}_m , \rho]  + \frac{1}{2}\gamma\bigl([B, \rho  B^{\dagger}] + [B \rho , B^{\dagger}]]\bigr)+ \frac{1}{2}\gamma e^{-\hbar\omega_0/k_BT}\bigl([B^{\dagger}, \rho B] + [B^{\dagger} \rho , B]\bigr) -\frac{1}{2}\Gamma [|1\rangle\langle 1|,[|1\rangle\langle 1|, \rho]]
\label{MME_T}
\end{eqnarray}
where the dissipation rate $\gamma = G_1(\omega_0)$, and the pure decoherence rate
\begin{equation}
\Gamma = 4{\xi_0}^2 G_1(0) + G_2(0).
\label{puredec}
\end{equation} 
\par
The master equation \eqref{MME_T} is a well known exactly solvable one which describes thermalization of the quantum harmonic oscillator
(with equilibrium position conditioned on the electronic state) accompanied by the independent decoherence process for a 2-level electron system. Any initial state evolves into the mixture of the \emph{conditioned Gibbs states}
\begin{equation}
\rho^{(0)} =  \bigl(1-e^{-\hbar\omega_0/k_B T}\bigr)|0\rangle\langle 0|\,e^{-\frac{\hbar\omega_0}{k_BT}A^{\dagger}A}, \quad
\rho^{(1)} = \bigl(1-e^{-\hbar\omega_0/k_B T}\bigr)|1\rangle\langle 1|\,e^{-\frac{\hbar\omega_0}{k_BT}(A^{\dagger}- \xi_0)(A - \xi_0)}
\label{gibbs_biased}
\end{equation} 
with weights given by the initial occupation probabilities of electronic states $|0\rangle , |1\rangle$.
\subsection{Coupling to exciton bath}
The interaction between the exciton bath and the molecule which is responsible for the process of ``charging"' of molecular battery can be derived from the fundamental (screened) Coulomb interaction between electrons in the whole system. Within the (nonrelativistic) formalism of the quantum field theory we describe the total electronic systems in terms of position-dependent (we omit spin variables for simplicity) fermionic quantum fields $\Psi(\mathbf{r})$, $\Psi^{\dagger}(\mathbf{r})$. Then the Coulomb interaction is described by the following Hamiltonian
\begin{equation}
H_{int} = \int\int \Psi^{\dagger}(\mathbf{r})\Psi^{\dagger}(\mathbf{r}') V(\mathbf{r},\mathbf{r}')\Psi(\mathbf{r})\Psi(\mathbf{r}')\,
d^3 \mathbf{r} d^3 \mathbf{r'} ,
\label{coulomb}
\end{equation}
with the effective electron-electron interaction potential $ V(\mathbf{r},\mathbf{r}')$.
\par
The electronic quantum field can be decomposed using the single-electron wave functions : $\phi^a_k (\mathbf{r})$, $\phi^b_{\ell} (\mathbf{r})$ corresponding to bands A and B, and  $\chi_0(\mathbf{r})$ , $\chi_1(\mathbf{r})$ to electronic states of the molecule. Then
\begin{equation}
\Psi(\mathbf{r}) = \sum_k \phi^a_k (\mathbf{r})\, a_k + \sum_{\ell} \phi^b_{\ell} (\mathbf{r})\, b_{\ell} +
\chi_0(\mathbf{r}) c_0 + \chi_1(\mathbf{r}) c_1
\label{qfield}
\end{equation}
where $c_0$ and $c_1$  are new fermionic annihilation and creation operators for an electron localized in the molecule.
\par
Inserting \eqref{qfield} into \eqref{coulomb} one obtains a sum of different  terms each containing two annihilation and two creation operators. Most of the terms describe the electron-electron interaction in the excitonic bath which were already taken into account in the self-consistent single-electron Hamiltonian \eqref{ham_electrons}, the other describe some residual potential between the bath and the molecule. The only terms which can account for the energy transfer between the bath and the molecule are represented by the following effective interaction Hamiltonian
\begin{equation}
H^{(3)}_{int} = \big(c_1^{\dagger} c_0 + c_0^{\dagger} c_1\big)\big[\sum_{k\ell} (g_{k\ell} a_k^{\dagger} b_{\ell} + \bar{g}_{k\ell} b_{\ell}^{\dagger}a_k)\big] , 
\label{coulomb1}
\end{equation}
where the coupling coefficients $g_{kl}$ might be in principle computed from the fundamental interaction Hamiltonian. The energy transfer is effective under the  condition $E_g \simeq E_{el}$  which implies that  only the terms containing the operators 
$a_k^{\dagger} b_{\ell}  c_1^{\dagger} c_0 $ and  $ b_{\ell}^{\dagger}a_k c_0^{\dagger} c_1$ describe real resonant exchange process depicted on Fig.1.
Notice that because we consider only a single electron occupying the molecule the operator $c_1^{\dagger} c_0 $ is equivalent to $|1\rangle\langle 0|$ and hence the Hamiltonians \eqref{coulomb1} is a special case of \eqref{Ham_int_3}.
\subsection{Charging the battery}
The interaction between the exciton bath and the molecular battery yields irreversible processes in the battery which can be described by the additional term added to the QMME \eqref{MME_T}, derived again using weak coupling formalism. The first step is the computation of the coupling spectrum given by the formula \eqref{spectral_bath_T} with $F_3 = \sum_{k\ell} g_{kl}(a_k^{\dagger} b_{\ell} + b_{\ell}^{\dagger}a_k)$ and for the thermal equilibrium state replaced by the stationary non-equilibrium one \eqref{grand}.
The standard computation yields the following results, for $\omega \geq 0$,
\begin{equation}
G_3(\omega)= \frac{1}{\hbar}\sum_{k\ell} |g_{k\ell}|^2 \delta\bigl(E_{a}(k) - E_{b}(\ell)-\hbar\omega \bigr) f_b(\ell) [1- f_a(k)]
\label{spectral_ex}
\end{equation}
and for negative frequencies
\begin{equation}
G_3(-\omega)= \frac{1}{\hbar}\sum_{k\ell} |g_{k\ell}|^2 \delta\bigl(E_{a}(k) - E_{b}(\ell)- \hbar\omega \bigr) f_a(k) [1- f_b(\ell)]
\label{spectral_ex1}
\end{equation}
with the Fermi-Dirac distributions \eqref{FD}.

Inserting \eqref{FD} into  \eqref{spectral_ex} and \eqref{spectral_ex1} yields the KMS relation
\begin{equation}
G_3(-\omega)= e^{-(\hbar\omega - \Delta\mu)/k_BT} G_3 (\omega).
\label{KMS_ex}
\end{equation}
which describes the ratio of transition probabilities per unit time of excitation and relaxation processes. This relation concerns a pair of energy levels of any system weakly  coupled to the excitonic bath.
\par
In order to discuss the shape of the coupling spectrum $G_3 (\omega)$ we need some estimations for the magnitudes of relevant parameters of the model. The typical parameters satisfy the conditions
\begin{equation} 
E_{el} \simeq E_g \simeq \Delta\mu\sim 1 eV >>  \hbar \omega_0 \sim 0.1 eV ,\quad k_B T \sim 0.01 eV .
\label{parameters}
\end{equation}
The existence of the gap and energy conservation hidden in Dirac deltas in \eqref{spectral_ex}, \eqref{spectral_ex1} imply that $G_3 (\omega)$ can be strictly positive only for Bohr frequencies 
$\omega > E_g$.
Notice that because the  states of the molecule which are coupled through the interaction with the exciton bath must involve different electronic states the set of relevant (and positive)  Bohr frequencies is the following
\begin{equation}
\{\omega\} = \{\omega_{el} +  m\omega_0 \geq 0 ;{m\in\mathbf{Z}}\}, 
\label{bohr}
\end{equation}
where $\omega_{el} = E_{el}/\hbar$.
\par
Any initial state of the battery weakly coupled to the excitonic  baths and thermalized by the mechanism described in the previous section tends to a final stationary state which according to  \eqref{gibbs_biased}, \eqref{KMS_ex} and  \eqref{bohr} has form
\begin{equation}
\rho_{st}=  \bigl[1+e^{-(E_{el} - \Delta\mu)/k_B T}\bigr]^{-1} \bigl( \rho^{(0)} + e^{-(E_{el} -  \Delta\mu)/k_B T} \rho^{(1)}\bigr),
\label{stat_battery}
\end{equation} 
and generally differs from the equilibrium state of battery given by \eqref{stat_battery} with $\Delta\mu =0$. 
\par
\textbf{Remark}
The computation of the coupling spectrum $G_3(\omega)$ was done using as dynamics of the exciton bath the unitary evolution governed by the Hamiltonian \eqref{ham_electrons} and not the full irreversible dynamics including the interaction with hot and cold baths. However, the dissipative effects caused by the later produce only a certain indeterminacy in the energy conservation which is expressed here by the Dirac deltas in formulas \eqref{spectral_ex}, \eqref{spectral_ex1}. The coupling spectrum $G_3(\omega)$ is essentially different from zero for the value $\omega > E_g/\hbar$ which are high in comparison to the relaxation rates induced by cold and hot bath (weak coupling assumption). Therefore, the dissipative effects lead only to replacement of Dirac deltas by their regularized models with the widths of the order of ($\hbar\times$ relaxation rates) what does not change essentially the coupling spectrum. 

\subsection{Work extraction from the battery}

The battery  in the stationary state \eqref{stat_battery}  is a source of useful work. The maximal extractable work is given by \emph{ergotropy}, the name introduced in \cite{ergo} and defined for an arbitrary state $\rho$ of a quantum system equipped with a Hamiltonian $H$ as 
\begin{equation} 
W_{max} = \mathrm{Tr}(\rho\, H) -  \mathrm{Tr}(\sigma_{\rho}\, H) ,
\label{ergo}
\end{equation}
where $\sigma_{\rho}$ is a unique \emph{passive state} \cite{Pusz} associated with $\rho$. This state is diagonal in the Hamiltonian basis, and possesses the same eigenvalues as $\rho$, ordered in the non-increasing order with energy.
\par
For a system composed of $n$ identical,  and non-interacting subsystems one can define the asymptotic maximal work, per subsystem, extractable by application of arbitrary many-particle external Hamiltonian perturbations
\begin{equation} 
\bar{W}_{max} = \lim_{n\to\infty} \frac{1}{n}\mathrm{Tr}\bigl(\otimes_n\rho\, H_n) -  \mathrm{Tr}(\sigma_{\otimes_n\rho}\, H_n\bigr) ,
\label{ergo1}
\end{equation}
where $H_n$ is a sum of identical single-subsystem Hamiltonians.
\par
It was shown in \cite{AlFan} that 
\begin{equation} 
\bar{W}_{max} = \mathrm{Tr}(\rho\, H) -  \mathrm{Tr}(\rho_{\bar{\beta}}\, H),
\label{ergo_free}
\end{equation}
where $\rho_{\bar{\beta}} = Z^{-1}e^{-\bar{\beta}H}$ is a unique Gibbs state with the same von Neumann entropy as the state $\rho$.
Generally, $\bar{W}_{max} \geq W_{max} $, is much easier to compute and the formula \eqref{ergo_free} can be interpreted in terms of non-equilibrium free energy.
\par
In the case of  molecular battery the basic ingredients of the relation \eqref{ergo_free} can be computed analytically. Namely, the von Neumann entropies of the stationary state \eqref{stat_battery} and the Gibbs state of the battery at the inverse temperature $\bar{\beta}$ read
\begin{eqnarray}
\label{entropy1}
S(\rho_{st}) =  \frac{\beta\hbar\omega_0}{e^{\beta\hbar\omega_0}-1} -\ln\bigl[1- e^{-\beta\hbar\omega_0} \bigr] + 
h\bigl[(1+ e^{-\beta\hbar(\Delta\mu - E_{el})})^{-1}\bigr],\\
\label{entropy2}
S(\rho^{(m)}_{\bar{\beta}}) =  \frac{\bar{\beta}\hbar\omega_0}{e^{\bar{\beta}\hbar\omega_0}-1} -\ln\bigl[1- e^{-\bar{\beta}\hbar\omega_0} \bigr] + 
h\bigl[(1+ e^{-\bar{\beta}\hbar E_{el}})^{-1}\bigr]  ,
 \\
h[x] = - x\ln x - (1-x) \ln(1-x), \quad x\in[0 , 1] .
\label{entropy3}
\end{eqnarray}
Now from the equation $S(\rho_{st})= S(\rho^{(m)}_{\bar{\beta}})$ one can numerically compute $\bar{\beta}$ and then insert into the final formula for the extractable work
\begin{equation} 
\bar{W}_{max} = \Bigl(\frac{\beta\hbar\omega_0}{e^{\beta\hbar\omega_0} -1} - \frac{\bar{\beta}\hbar\omega_0}{e^{\bar{\beta}\hbar\omega_0} -1}\Bigr) + E_{el} \Bigl(\frac{1}{e^{\beta(E_{el} - \Delta\mu)} + 1} - \frac{1}{e^{\bar{\beta}\omega_0} + 1} \Bigr).
\label{ergo_free1}
\end{equation}
For low temperature regime one can obtain from \eqref{ergo_free}
\begin{equation} 
\bar{W}_{max}(T=0) = E_{el}\, \Theta(\Delta\mu - E_{el}),
\label{ergo_free2}
\end{equation}
where $\Theta(x)$ is the Heaviside function.

\subsection{Stability of molecular battery}

The molecular battery in the excited electronic state $\rho^{(1)}$ detached from the excitonic bath still remains in contact with the heat bath at the ambient temperature $T$ corresponding to various environmental degrees of freedom. Such interaction leads to spontaneous discharging of the battery as well as to much less probable (suppressed by the Boltzmann factor $e^{-E_{el}/k_B T}$) spontaneous recharging. Of course, the lower is the discharging rate the more useful is the battery. 

The dynamics of the molecule in the environment at ambient temperature is given by the Master equation of the form
\begin{equation} 
\frac{d}{dt} \rho = -\frac{i}{\hbar}[H_{m} , \rho] + \mathcal{L}_0  \rho +\mathcal{L}_1  \rho .
\label{MMEbat}
\end{equation}
Here, $\mathcal{L}_0$ describes  thermalization process of the reaction coordinate and pure decoherence of the electronic degree of freedom 
given by the non-Hamiltonian part of the RHS of eq.\eqref{MME_T}. The generator $\mathcal{L}_1$ accounts for transitions between different electronic states caused by thermal fluctuations and derived using the interaction Hamiltonian of the type \eqref{Ham_int_3}. This  additive structure can be justified assuming that the leading interaction of the molecule with an environment is mediated by the dipole moment of the molecule which is a sum of contributions from nuclei and electrons. The first one enters the interaction Hamiltonian of the form \eqref {Ham_int_1} leading to reaction coordinate thermalization and the second one leads to \eqref{Ham_int_3} and accounts for spontaneous discharging/charging processes.
\par
As we shall see the structure of $\mathcal{L}_1$ is quite involved and therefore we cannot solve analytically the full QMME \eqref{MMEbat}. However, the stability of battery can be characterized by the initial rate of the probability flow out of the ``excited Gibbs state" $\rho^{(1)}$  (see \eqref{gibbs_biased}) 
\begin{equation}
\Gamma_{\downarrow} = \mathrm{Tr}\Bigl(|0\rangle\langle 0|\mathcal{L}_1\rho^{(1)} \Bigr).
\label{tunrate_A}
\end{equation}
Indeed, the state $\rho^{(1)}$ commutes with $H_m$ and is stationary with respect to $\mathcal{L}_0 $ hence only $\mathcal{L}_1$ contributes to the initial dynamics of $\rho^{(1)}$.

The generator $\mathcal{L}_1$ can be obtained again using the weak coupling approach and possesses the following form (we use a different ordering of terms than in the formula \eqref{MME_T})
\begin{eqnarray}
\label{L_1}
\mathcal{L}_1\rho &=& \sum_{m\in\mathbf{Z}}G(\omega_{el}+ m\omega_0 )\Bigl\{\bigl[({V}_m  \rho {V}^{\dagger}_m 
-\frac{1}{2}\{{V}^{\dagger}_m {V}_m ,\rho \} \bigr]  \\
\nonumber
&+& e^{-(E_{el}+ \hbar m\omega_0 )/k_BT}\bigl[({V}^{\dagger}_m  \rho {V}_m 
-\frac{1}{2}\{{V}_{m}{V}^{\dagger}_m ,\rho \} \bigr] \Bigr\}
\end{eqnarray}
The transition operators ${V}_m$ can be computed  by decomposition of of \eqref{Fourier_2} into terms oscillating with the frequencies $-(E_{el}/\hbar + m\omega_0)$ and using \eqref{weyl} what gives
\begin{equation}
V_m = e^{-|\xi_0|^2/2} \Bigl[\sum_{k\geq \max\{0, -m\}} \frac{(-1)^k {\xi_0}^{(2k+m)}}{k!(k+m)!} \bigl(A^{\dagger}\bigr)^{k+m} A^k\Bigl] W(-\xi_0) |0\rangle\langle 1| .
\label{opV}
\end{equation}
Combining the eqs. \eqref{gibbs_biased}, \eqref{tunrate_A},  \eqref{L_1} and \eqref{opV} one obtains
\begin{equation}
\Gamma_{\downarrow} = \sum_{m\in\mathbf{Z}}G(\omega_{el}+ m\omega_0 )\mathrm{Tr}\bigl({V}_m  \rho^{(1)} {V}^{\dagger}_m \bigr) = 
 = e^{-S} \bigl(1-e^{-\hbar\omega_0/k_B T}\bigr)\sum_{m\in\mathbf{Z}}G(\omega_{el}+ m\omega_0 )\mathrm{Tr}\bigl(e^{-\frac{\hbar\omega_0}{k_BT}A^{\dagger}A} v^{\dagger}_m v_m\bigr)
\label{rate}
\end{equation} 
where
\begin{equation}
v_m = \sum_{k\geq \max\{0, -m\}} \frac{(-1)^k {\xi_0}^{(2k+m)}}{k!(k+m)!} \bigl(A^{\dagger}\bigr)^{k+m} A^k  .
\label{opv}
\end{equation}
As typically $\hbar \omega_{el} >> \hbar\omega_0 >> k_B T$ it is instructive to compute $\Gamma_{\downarrow}$ in the limit case $T =0$. Then
\begin{equation}
\Gamma_{\downarrow}(T=0)= e^{-S} \sum_{m\in\mathbf{Z}}G(\omega_{el}+ m\omega_0 )\langle 0| v^{\dagger}_m v_m |0\rangle = \sum_{m = 0}^{\infty}G(\omega_{el}- m\omega_0 ) e^{-S} \frac{S^m}{m!}.
\label{rate0}
\end{equation} 
For $S >> 1$ the Poisson distribution on the RHS of \eqref{rate0} is concentrated around the mean value $\bar{m} = S$ what, for smooth enough coupling spectrum $G(\omega)$ leads to the final formula
\begin{equation}
\Gamma_{\downarrow}(T=0) \simeq G(\omega_{el}- S\omega_0 ) .
\label{rate01}
\end{equation} 
Due to the KMS condition \eqref{KMS},  for $T\to 0$, the coupling spectrum  $G(\omega)$ vanishes for $\omega < 0$. Therefore, the charged state of the molecular battery becomes stable for $S > \frac{\omega_{el}}{\omega_0}$. One can expect that in the case of finite temperatures the rapid suppression of spontaneous discharging is replaced by the Boltzmann factor $\sim e^{-(S\hbar\omega_0 - E_{el})}$ (compare with the computation in \cite{Alicki:2013} performed for the model with $E_{el}=0$).

One should notice that the strategy of increasing Huang-Rhys factor to improve the performance of battery  possesses its limitations. Namely, the Huang-Rhys factor appears also in the decoherence rate \eqref{puredec} in the QMME \eqref{MME_T}. To extract work from the battery one has to apply external time-dependent perturbation of the Hamiltonian which does not commute with the Hamiltonian of the battery. Under such perturbation, the previous pure decoherence process causes now energy dissipation reducing the efficiency of work extraction.

\section{Concluding remarks}

Although quantum theory of open systems is an active research topic in  the field of chemical, biological, and condensed-phase physics, in most cases little attention is payed to the mathematical consistency of the applied formalism. The typical examples are the various versions of Redfield equation or approximations involving path integrals which neither satisfy the requirements of transforming density matrices into density matrices, nor the general restrictions imposed by thermodynamics. The presented method based on systematic and careful weak coupling approximation allows  to construct the evolution equations of the GKLS type satisfying the mentioned above natural requirements. Moreover,  those equations possess a highly symmetric mathematical structure which allows to obtain analytical results for quite complex systems as illustrated by the model of molecular battery charged by excitons. Those analytical results can be extrapolated to more complicated situations which require numerical methods.  
\par
Quite often it is argued that the QMME yield much less accurate solutions than, for instance, Redfield equation, in particular for the short-time evolution \cite{Redfield}. In the Appendix the basis approximations, used  in the derivation of QMME, are discussed. It is shown, indeed, that a clear separation of various time scales is necessary to justify the Markovian approximation. Generically, solutions of QMME describe accurately the evolution of open systems for long enough times what is often sufficient to discuss the approach to asymptotic  stationary state. In the case of various not well-separated time scales a careful choice of the reference physical Hamiltonian used in the first step of the derivation -- transition to the interaction picture -- can improve the accuracy of QMME (see e.g. \cite{Vsystem}).
Finally, the ``refined weak coupling" formalism \cite{Alicki:89}, \cite{Rivas} allows to construct  completely positive dynamical maps interpolating with a good accuracy between the strongly non-Markovian short-time dynamics and asymptotic long-time evolution governed by QMME.

\section*{Acknowledgement} 
The work is a part of the ICTQTIRAP project of FNP, financed by structural funds of EU.

\section*{Appendix. Quantum Markovian Master Equations}

The brief derivation of QMME based on cumulant expansion and refined weak coupling approach is presented and basic properties of QMME
are discussed, in particular for heat and chemical baths \cite{Alicki:89}\cite{Rivas}.

\subsection{From refined weak coupling to MME}

A  quantum system  with a "bare" system Hamiltonians $H^{0}$ interacts with a reservoir
equipped with the  Hamiltonian $H_{R}$, by means of  the interaction Hamiltonian $ H_{int}=\lambda S\otimes R$ with $S =S^{\dagger}$,
$R =R^{\dagger}$, with $\lambda $ being a ``small" coupling strength ( generalization to more complicated $H_{int}$ is straightforward).
The initial state of the baths satisfies
\begin{equation}
[\rho_R , H_R] = 0,\ \mathrm{Tr}(\rho_R\, R) =0 .
\label{ass}
\end{equation}
\par
The reduced, system-only dynamics in the interaction picture is defined as a partial trace
\begin{equation}
\rho (t)=\Lambda (t)\rho \equiv \mathrm{Tr}_R \bigl(U_{\lambda}(t)\rho\otimes\rho_R U_{\lambda}(t)^{\dagger}\bigr)
\label{red_dyn}
\end{equation}
where the unitary propagator in the interaction picture is given by the ordered exponential
\begin{equation}
U_{\lambda}(t) = \mathcal{T}\exp\Bigl\{\frac{-i\lambda}{\hbar}\int_0^t S(s)\otimes R(s)\,ds\Bigr\}
\label{prop_int}
\end{equation}
with
\begin{equation}
S(t) = e^{(i/\hbar)Ht} S e^{(i/\hbar)Ht} ,\  R(t)= e^{(i/\hbar)H_R t} R e^{-(i/\hbar)H_R t}.
\label{prop_int1}
\end{equation}
Notice, that $S(t)$ is defined  with respect to the renormalized, \emph{%
physical} Hamiltonian $H$  which can be expressed as
\begin{equation}
H=H^{0}+\lambda ^{2}H_{1}^{\mathrm{corr}}+\cdots .  
\label{eq:H_S}
\end{equation}
The renormalizing counterterms containing powers of $\lambda $ are
often called \emph{ Lamb-shift} corrections and are due to the
interaction with the bath.
\par
The cumulant expansion for the reduced dynamics can be defined as
\begin{equation}
\Lambda (t)=\exp \sum_{n=1}^{\infty }[\lambda ^{n}\mathcal{K}^{(n)}(t)].
\label{cumulant}
\end{equation}
Comparing terms with the same power of $\lambda$ in \eqref{cumulant} and in the Dyson expansion of \eqref{red_dyn} one finds that $\mathcal{K}^{(1)}=0$ and the Born approximation to the exact dynamics in the interaction picture has form: 
\begin{equation}
\Lambda_B (t)=\exp [\lambda ^{2}\mathcal{K}^{(2)}(t)],
\label{Born}
\end{equation}
with
\begin{equation}
\mathcal{K}^{(2)}(t)\rho =\frac{1}{\hbar^2}\int_{0}^{t}ds\int_{0}^{t}du\,\Bigl[ F(s-u) \bigl(S(s)\rho S(u) - \frac{1}{2}\{S(u)S(s), \rho  \}\bigr) \Bigr]
- \frac{i}{\hbar}[H_{L}(t), \rho] .
\label{eq:K(t)}
\end{equation}
Here $F(s)= \mathrm{Tr}(\rho _{R}R(s)R)$ and the Hamiltonian term
\begin{equation}
H_{L}(t) = i\frac{1}{2\hbar}\int_{0}^{t}ds\int_{0}^{t}du\, F(s-u) [S(s) ,S(u) ]
\label{eq:K(t)1}
\end{equation}
is often called Lamb-shift Hamiltonian. In the Schroedinger picture and for the Markovian limit $H_{L}(t)$ can be omitted because its contribution is compensated by the counterterm in the renormalized Hamiltonian \eqref{eq:H_S}.
\par
Notice, that the mathematical structure of the superoperator $\mathcal{K}^{(2)}(t)$ allows to recast it into the standard Gorini-Kossakowski-Lindblad-Sudarshan (GKLS) form and hence its exponential \eqref{Born} defines a family of completely positive trace preserving dynamical maps.
This mathematically consistent approximation was proposed first in \cite{Alicki:89} and  rediscovered recently in \cite{Rivas} where  numerical tests shown its usefulness.
\par
We show now that for long enough time the following Markovian approximation is valid
\begin{equation}
\lambda^2\mathcal{K}^{(2)}(t)\simeq t\mathcal{L}  \label{eq:L}
\end{equation}
where $\mathcal{L}$ is a GKLS generator. To find its form we first decompose
$S(t)$ into its Fourier components
\begin{equation}
S(t)=\sum_{\{\omega\} } e^{-i\omega t}S_{\omega }, \quad S_{-\omega }= S_{\omega }^{\dagger}
\label{eq:S}
\end{equation}
where the set $\{\omega\}$ contains \emph{Bohr frequencies} of the physical Hamiltonian with the spectral decomposition
\begin{equation}
H= \sum_k \epsilon_k |k\rangle\langle k|,\  \omega = \epsilon_k - \epsilon_l .  
\label{Bohr}
\end{equation}
Then we can rewrite the expression (\ref{eq:K(t)}) (with neglected Hamiltonian term) as
\begin{equation}
\mathcal{K}^{(2)}(t)\rho =\frac{1}{\hbar^2}\sum_{\omega ,\omega ^{\prime }}S_{\omega }\rho S_{\omega
^{\prime }}^{\dag }\int_{0}^{t}e^{-i(\omega -\omega ^{\prime
})u}du\int_{-u}^{t-u}F(\tau )e^{-i\omega \tau }d\tau +(\mathrm{similar}\text{ 
}\mathrm{terms}).  
\label{eq:K2}
\end{equation}
and use two crucial approximations:
\begin{equation}
\int_{0}^{t}e^{-i(\omega -\omega ^{\prime })u}du\approx t\delta _{\omega
\omega ^{\prime }}, \  \int_{-u}^{t-u}F(\tau )e^{-i\omega \tau }d\tau \approx {G}(\omega
)=\int_{-\infty }^{\infty }F(\tau )e^{-i\omega \tau }d\tau \geq 0 ,
\label{eq:rep1}
\end{equation}
which are valid for $t\gg \max \{1/(\omega -\omega ^{\prime })\}$. Applying these two approximation we obtain $K(t)\rho _{S}=(t/\hbar^2)\sum_{\omega
}S_{\omega }\rho _{S}S_{\omega }^{\dag }{G}(\omega )+(\mathrm{similar}$ $%
\mathrm{terms})$, and hence it follows from Eq.~(\ref{eq:L}) that $\mathcal{L}$ is a special case of the GKLS generator derived rigorously for the first time by Davies \cite{Davies}. Returning to the Schroedinger picture one obtains the following QMME:
\begin{eqnarray}
\frac{d\rho }{dt} &=&-\frac{i}{\hbar}[H,\rho ]+\mathcal{L}\rho ,  \notag \\
\mathcal{L}\rho  &\equiv &\frac{\lambda ^{2}}{2\hbar^2}\sum_{\{\omega \}}G(\omega
)([S_{\omega },\rho S_{\omega }^{\dagger }]+[S_{\omega }\rho ,S_{\omega
}^{\dagger }])  
\label{Dav}
\end{eqnarray}
\textbf{Remarks}:

\noindent (i) The absence of off-resonant terms in Eq.~(\ref{Dav}) (so-called secular approximation) is the crucial property of the Davies generator which can be 
interpreted as averaging in time of fast oscillating terms. It implies also the commutation of $\mathcal{L}$
with the Hamiltonian part $[H ,\cdot]$. In the case of almost degenerated Bohr frequencies one can use in the interaction picture
a modified Hamiltonian with certain strictly degenerated Bohr frequencies to improve the approximation \cite{Vsystem}.

\noindent (ii) The positivity  $G(\omega )\geq 0$  follows from the Bochner's theorem and is a necessary condition for
the complete positivity of the QMME.

\noindent (iii) The presented derivation shows implicitly that the notion of
\emph{bath's correlation time}, often used in the literature, is generally not
well-defined -- Markovian behavior involves a rather complicated cooperation
between system and bath dynamics.  
\par
It follows from (iii) that, contrary to what is often done in
phenomenological treatments, \emph{one cannot combine arbitrary }$H$%
\emph{'s with a given GKLS generator}. This is particularly important
in the context of thermodynamics of controlled quantum open system, where it is common to fix Markovian, irreversible part of
dynamics and apply arbitrary control Hamiltonians. Uncontrolled derivations of QMMEs can easily lead to violation of the laws of thermodynamics \cite{Benatti} \cite{Levy},\cite{Hofer}.
\par
\textbf{Remark} All the  results including the expressions for \eqref{eq:K(t)} and \eqref{Dav} can be easily extended to  more complicated
interaction Hamiltonians
\begin{equation}
H_{int}= \lambda\sum_{\alpha} S_{\alpha}\otimes R_{\alpha} , \, \mathrm{or} \quad
H_{int}= \lambda\sum_{\alpha} \bigl(S_{\alpha}\otimes R^{\dagger}_{\alpha} + S^{\dagger}_{\alpha}\otimes R_{\alpha}\bigr) .
\label{Int_app}
\end{equation}
In particular the most general form of QMME obtained by the weak coupling procedure reads
\begin{equation}
\frac{d\rho }{dt} =-\frac{i}{\hbar}[H,\rho ]+
\frac{\lambda ^{2}}{2\hbar^2}\sum_{\alpha\beta}\sum_{\{\omega \}}G_{\beta\alpha}(\omega
)([S_{\alpha}(\omega ),\rho S_{\beta}(\omega )^{\dagger }]+[S_{\alpha}(\omega )\rho , S_{\beta}(\omega )^{\dagger }]) , 
\label{Dav_gen}
\end{equation}
where for all $\{\omega \}$,  the matrix $[G_{\beta\alpha}(\omega)]$ is positively defined.

)
\subsection{Thermal and chemical reservoirs}

If the reservoir is a quantum system at thermal equilibrium state the additional Kubo-Martin-Schwinger (KMS) condition holds
\begin{equation}
G(-\omega) = \exp\Bigl(-\frac{\hbar\omega}{k_B T}\Bigr) G(\omega),  
\label{KMS_app}
\end{equation}
where $T$ is the bath's temperature. As a consequence of (\ref{KMS_app}) the Gibbs state 
\begin{equation}
\rho_{\beta} = Z^{-1} e^{-\beta H}, \ \beta= \frac{1}{k_B T}  
\label{gibbs_app}
\end{equation}
is a stationary solution of (\ref{Dav}). Under mild conditions (e.g : "the only system operators commuting with $H$ and $S$ are scalars") the Gibbs state is a unique stationary state and any initial state relaxes towards equilibrium ("0-th law of thermodynamics"). A convenient parametrization of the corresponding \emph{thermal generator} reads
\begin{equation} 
\mathcal{L}\rho  = \frac{1}{2}\sum_{\{\omega\geq 0\} }\gamma(\omega)\bigl\{([S_{\omega },\rho S_{\omega }^{\dagger }]+[S_{\omega }\rho ,S_{\omega}^{\dagger }]) + e^{-\hbar\beta\omega}([S_{\omega }^{\dagger},\rho S_{\omega }] +[S_{\omega }^{\dagger }\rho ,S_{\omega}])\bigr\} 
\label{Dav_therm}
\end{equation}
where 
\begin{equation}
\gamma(\omega)= \frac{\lambda^2}{\hbar^2} \int_{-\infty}^{+\infty}e^{-i\omega t} \mathrm{Tr}\bigl(\rho_R\, e^{iH_R t/\hbar}\,R\, e^{-iH_R t/\hbar}R\bigr)\, dt .  
\label{relaxation}
\end{equation}
A \emph{chemical bath} \cite{Universal} is a collection of $K$ separated baths each consisting of many molecules of a given type. The whole bath  is described by the density matrix corresponding to the grand canonical ensemble at the temperature  $T_1$ and the chemical potentials $\mu_j$
\begin{equation}
\tilde{\rho}_R = Z^{-1} \exp\Bigl\{-\frac{1}{T_1} \bigl(H_R - \sum_{j=1}^K \mu_j N_j\bigr)\Bigr\}.
\label{grand_app}
\end{equation}
Here, $N_j$ is the operator describing the number of $j$-type molecules in the bath and $H_R$ is the total Hamiltonian of the chemical bath satisfying 
\begin{equation}
[N_j , H_R ] = 0 , \mathrm{for\ all} \,  j=1, 2,...,K .
\label{grand1}
\end{equation}
The condition \eqref{grand1} means that the  molecules in the chemical bath do not undergo chemical reactions unless they became coupled by the quantum  system. This coupling can be  described by the rotating-wave approximation version of the interaction Hamiltonian \eqref{Int_app}
\begin{equation}
H_{int}= \lambda \bigl(S^{-}\otimes R^{+}  + S^{+}\otimes R^{-}\bigr) , \quad  S^{+} =  (S^{-})^{\dagger} , R^{+} =  (R^{-})^{\dagger} .
\label{ham_RW}
\end{equation}
Here,  $S^{+}$  increases the energy of the system and the operator $R^{-}$ describes the possible transitions from higher to lower  energy levels of the bath including  chemical reactions. One can think about 
$R^{-}$ as a sum of products  of  operators describing three possible processes: a) transitions preserving all numbers of molecules, b) ``annihilation operators" corresponding to reactants, c) ``creation operators" corresponding to reaction products. Hence, 
$R^{-}$ satisfies the following relation
\begin{equation}
[\sum_{j=1}^K \mu_j N_j ,  R^{-}] = \bigl(\sum_{j=1}^K \nu_j\mu_j\bigr)  R^{-} ,
\label{stoichio}
\end{equation}
where $\nu_j$ are \emph{stoichiometric coefficients}, negative for reactants and positive for reaction products. The expression 
\begin{equation}
\Delta g =  -\sum_{j=1}^K \nu_j\mu_j 
\label{free_gibbs_app}
\end{equation}
can be interpreted as the Gibbs free energy transferred during a single step of chemical reaction to the system. Obviously, the operator  $ S^{-}\otimes R^{+}$ describes the reverse processes consuming energy from the system. If the operators $R^{\pm}$ commute with all number operators $N_j$, then $\Delta g = 0$, no chemical reactions occur, and the bath acts as a purely thermal bath at the temperature $T_1$.
\par
Applying again the standard derivation of the QMME one obtains the equation \eqref{Dav_therm} with the modified relaxation coefficient 
\begin{equation}
\label{coupling_spectrum1}
\gamma_{\downarrow} (\omega) = \frac{\lambda^2}{\hbar^2}\int_{-\infty}^{\infty}e^{-i\omega t}\,\mathrm{Tr}\bigl(\tilde{\rho}_R\, R^{-}(t)R^{+}\bigr) dt ,
\end{equation}
and $e^{-\beta\omega}\gamma (\omega)$ replaced by the excitation coefficient 
\begin{equation}
\label{coupling_spectrum2}
\gamma_{\uparrow} (\omega) = \frac{\lambda^2}{\hbar^2}\int_{-\infty}^{\infty}e^{-i\omega t}\,\mathrm{Tr}\bigl(\tilde{\rho}_R\, R^{+}(t)R^{-}\bigr) dt .
\end{equation}
The fictitious dynamics governed by the  \emph{extended Hamiltonian} $\tilde{H}_R = H_R - \sum_{j=1}^K \mu_j N_j$ and applied to the operators $R^{\pm}$ yields
\begin{equation}
\label{extended}
\tilde{R}^{\pm}(t)= e^{i\tilde{H}_R t} R^{\pm} e^{-i\tilde{H}_R t} =  R^{\pm}(t) e^{\mp i\Delta g\, t}.
\end{equation}
Using \eqref{extended},  and again the KMS condition for  the grand canonical ensemble $\tilde{\rho}_R$ treated as an ``extended" Gibbs state and  the ``extended" operators $\tilde{R}^{\pm}(t)$ one obtains the following relation 
\begin{equation}
\label{KMS1}
\gamma_{\uparrow} (\omega) = \exp\Bigl\{-\frac{1}{T_1} (\omega - \Delta g )\Bigr\}\,\gamma_{\downarrow} (\omega)  
\end{equation}
which generalizes  the detailed balance condition \eqref{KMS_app} to chemical baths. One can show that such structure of  QMME implies the validity of the second law of thermodynamics \cite{SpohnLeb} - \cite{QT-review} and Onsager relations in the close to equilibrium regime\cite{Alicki:76}.

\end{document}